\documentclass[twocolumn,floatfix,prb,aps,superscriptaddress,10pt,letterpaper]{revtex4}
\usepackage{amsfonts}
\usepackage{amsmath}
\usepackage{graphicx}
\allowdisplaybreaks[1]%biutiful equation breaker!!!
\usepackage{braket}
\usepackage{bm}
\usepackage[backref,pdffitwindow,colorlinks,linkcolor={blue},citecolor={red}]{hyperref}
\newcommand{\hl}[1]{{\color{blue} #1}}
\begin{document}

\title{Ab Initio Theory of the Drude Plasma Frequency}
\author{Bernardo S. Mendoza}\email{bms@cio.mx}
    \affiliation{Centro de Investigaciones en \'Optica,
                Le\'on, Guanajuato, M\'exico}
\author{W. Luis Moch\'an}
\affiliation{Instituto de Ciencias F{\'i}sicas, Universidad Nacional
 Aut\'onoma de M\'exico, Av. Universidad s/n, Col. Chamilpa, 62210
 Cuernavaca, Morelos, M\'exico.}

\begin{abstract}
We derive a theoretical expression to calculate the Drude plasma
frequency $\omega_D$ based on
quantum mechanical time dependent perturbation theory in the
long-wavelength regime.
We show that in general $\omega_D^2$ should be replaced by a
second rank tensor, the Drude tensor $\bm{\mathcal D}$, which
we relate to the integral over the Fermi surface of the convective
momentum flux tensor divided by
the magnitude of the Fermi velocity, and which is amiable to analytical and numerical
evaluation. We also obtain an
expression in terms of the average inverse mass tensor.
For the Sommerfeld's model of metals our  expression yields the ubiquitous
plasma frequency
$\omega^2_D=4\pi n_e e^2/m_e$.
We  compare our expressions to those of other previous theories.
The Drude tensor  takes into account the geometry of the unit cell and
may be calculated from first principles for isotropic as well as anisotropic metallic
systems. We present results for the noble metals, Ag, Cu, and Au
without stress and subject to isotropic and uniaxial strains, and we
compare the results to those available from experiment. We show that
within density functional theory, nonlocal potentials are necessary to
obtain an accurate Drude tensor.
\end{abstract}

\maketitle
%\tableofcontents

%%%%%%%%%%%%%%%%%%%%%%%%%%%%%%%%%%%%%%%

\section{Introduction}\label{intro}
The Drude theory for the electrical conductivity and the dielectric
response of metals has been well known for over a
century.\cite{drudeANP19001, drudeANP19002} Originally
based on the classical kinetic theory of gases, its results survived
Sommerfeld's reformulation in terms of a quantum Fermion gas. They
may also be obtained from the local limit of the non-local
longitudinal Lindhard dielectric function
\cite{ashcroft_solid_1976}. The purpose of this paper is to generalize
the Drude model using a fully quantum mechanical theory for metals
and to obtain local expressions that can be directly evaluated for
both isotropic and anisotropic systems starting from the Hamiltonian
of the system, which may contain many body and non-local interactions.

The Drude theory of metals, put forward back in 1900,\cite{drudeANP19001, drudeANP19002}
includes among its successes  a simple explanation that allows for
approximate estimates of properties of metals  whose full comprehension
required the development of the quantum theory of condensed
matter. Drude's theory of electrical and thermal conductivity was
based on the kinetic theory of gases applied to the conduction electrons
of the system, while the the core electrons
strongly bonded to the atomic nuclei are taken as an inert
entity. The precise assumptions of the Drude model are clearly
explained in textbooks of solid state theory like Ref.~\onlinecite{ashcroft_solid_1976}.
Drude assumed that
the electronic velocity distribution is given by the Maxwell-Boltzmann
distribution function, which leads to the wrong result for the specific
heat of metals of $3k_B/2$, where $k_b$ is Boltzmann's constant. It
was Sommerfeld, almost 25 years after the publication of Drude's results, that
recognized the fact that Fermi-Dirac (FD) statistics were required for
treating the electron correctly.  As it turns out, Sommerfeld's model
is essentially the classical electron gas model used by Drude, but using the FD
distribution for the valence electrons. Albeit, one of the first
success of Sommerfeld was to obtain the experimental linear behavior of the specific heat of
metals.

The  Drude model yields the following well known
result for the dielectric function $\epsilon(\omega)$ of a metal,\cite{ashcroft_solid_1976}
\begin{align}\label{ds.1}
\epsilon(\omega)=1 -\frac{\omega^2_D}{\omega^2}
,
\end{align}
where
\begin{align}\label{ds.2}
\omega_D=\left(\frac{4\pi n_e e^2}{m_e}\right)^{1/2}
,
\end{align}
is known as the Drude frequency,
where $n_e$, $-e$ and $m_e$ are the number density, charge and free mass of the
conduction electrons, and $\omega$ is the frequency of the light that perturbs
the electrons. \hl{In the Drude model $\omega_D$ corresponds also to the
frequency of the collective plasma oscillations of the system and is
usually denoted by $\omega_p$ and called the plasma frequency, though
this interpretation does not hold in the presence of further screening
mechanisims.}
Although Eq. \eqref{ds.1} gives a precise
description of the interaction of light with metals for photons with energy
below the threshold of electronic interband transitions,
a fully
quantum-mechanical derivation
would justify its successes.

In this article, we derive a closed expression for $\omega_D$ based on
quantum mechanical time dependent perturbation theory.
We show that in general $\omega_D^2$ should be replaced by a
second rank tensor, that we call the Drude tensor $\bm{\mathcal D}$, for which
we
derive a closed expression that is amiable to analytic evaluation in
the case of free independent electrons, and to numerical
evaluation for {\em ab initio} quantum mechanical calculations.
The Drude tensor  takes into account the geometry of the unit cell and
may be calculated from first principles for isotropic as well as anisotropic metallic
systems.

The article is
organized as follows. In Sec.~\ref{pa}, we present the
theoretical formalism, showing the main expressions used to
calculate the Drude tensor. In Sec.~\ref{im}
we derive $\omega_D$ within the Drude-Sommerfeld model as a special
case of our formalism, and
in Sec.~\ref{si} we present the procedure to numerically evaluate the
Drude tensor. In Sec.~\ref{res} we present results for the
noble metals, Ag, Cu and Au without stress and subject to an isotropic and
to an uniaxial strain, and we summarize our findings in Sec.~\ref{conc}.

\section{Analytic expression for the Drude plasma frequency}\label{pa}

In order to derive the analytic expression for the Drude plasma frequency,
we assume the electrons may be described through an independent
particle approximation, although we do allow for many-body effects
through an effective Hamiltonian that depends on all occupied states,
as in density functional theory.  The electrons interact with an
electromagnetic field which we assume is a classical field. Thus we describe quantum mechanical
matter interacting with classical fields. We neglect local field and
excitonic effects.\cite{onidaRMP02}
We write the one electron Hamiltonian
\begin{equation}
\hat H(t) = \hat H_{0} + \hat H_{I}(t),
\label{ache}
\end{equation}
as the sum of an unperturbed effective time-independent Hamiltonian $\hat H_{0}$
that describes the interaction of an electron with the crystalline
lattice and its effective interaction with the other
electrons, and an {\em interaction} Hamiltonian $\hat H_{I}(t)$ that
describes the interaction of the electron with a  time-dependent electromagnetic field.
We describe the
{\em state} of the system through
the one electron density operator $\hat{\rho}$, with which we can calculate the
expectation value of any single-particle observable $\hat{\mathcal{O}}$ as
$\braket{\hat{\mathcal O}} = \mbox{Tr}(\hat{\rho}\hat{\mathcal O})$
with Tr denoting the trace.
Within the {\em interaction picture} the density operator evolves in
time due to the interaction Hamiltonian according to
\begin{equation}
  \label{eq:Sch}
  i\hbar\frac{d}{dt}\hat{\rho}(t) = [\hat{H}_I(t), \hat{\rho}(t)],
\end{equation}
while the operators that correspond to all observables evolve through $\hat{H}_0$ according to
\begin{equation}
  \label{eq:Ovst}
  \hat{\mathcal{O}}(t) = \hat U^\dagger(t)\hat {\mathcal O}_s(t)\hat U(t),
\end{equation}
where $\hat {\mathcal O}_s(t)$ is the same observable in the Schrödinger
picture, given by $\hat{\mathcal O}(0)$ for operators that do not
depend {\em explicitly} on time, and
\begin{equation}
  \label{eq:U0}
  \hat U(t)=\exp({-i\hat H_{0}t/\hbar})
\end{equation}
is the non-perturbed unitary {\em time-evolution} operator.
Assuming the field is turned on adiabatically, we may integrate
\eqref{eq:Sch} to yield
\begin{equation}
  \hat{\rho}(t) =
  \hat{\rho}_{0} +
  \frac{1}{i\hbar}  \int_{-\infty}^{t}dt'
    [\hat{H}_{I}(t'),\hat{\rho}(t')],
\label{trans}
\end{equation}
where $\hat{\rho}_0$ is the
unperturbed, time-independent equilibrium
density matrix. We look for the standard perturbation series solution,
$\hat{\rho}(t) = \hat{\rho}_0 + \hat{\rho}^{(1)}(t) +
\hat{\rho}^{(2)}(t) + \ldots$, where the superscript denotes the order (power)
with which each term depends on the perturbation $\hat H_{I}(t)$.
Since we are interested only in the linear response, we concentrate our
attention on the 1-st order
term
\begin{equation}
  \label{eq:first}
  \hat\rho^{(1)}(t)=\frac{1}{i\hbar}\int_{-\infty}^t dt'[\hat H_I(t'),\hat \rho_0].
\end{equation}
We will take our system as a solid described by a non-perturbed
periodic Hamiltonian, whose eigenfunctions are Bloch states, $\ket{m\bm{k}}$,
characterized by a band index $m$ and a crystal momentum $\bm k$. For
$\hat H_I(t)$ we take an interaction with an electromagnetic
field with a wavelength much larger than the crystal parameter. Thus,
electronic transitions due to this interaction are vertical, i.e., they
conserve $\bm k$.

Within the dipole approximation, the
interaction Hamiltonian in the Length Gauge is given by
\cite{andersonPRB15}
\begin{equation}
\hat H_I(t)=e\hat{\bm r}(t)\cdot \bm E(t),
\label{rde}
\end{equation}
where $\hat{\bm r}(t)=\hat U^\dagger_0(t)\hat{\bm r}(0)U(t)$
is the
position operator of the electron at time $t$
 and $\bm E(t)$
the time dependent perturbing classical electric field.

From \eqref{eq:first} we obtain the first order density matrix
elements between Bloch states
\begin{widetext}
\begin{align}
  \rho^{(1)}_{nm}(\bm k;t)
  \equiv
  \braket{n\bm k|\hat{\rho}^{(1)}(t)|m\bm k}
&=    \frac{1}{i\hbar}\int_{-\infty }^{t}dt'
\braket{n\bm k|[\hat{H}_{I}(t'), \hat{\rho}_0]|m\bm k}
\nonumber\\
&=    -\frac{ie}{\hbar}\int_{-\infty }^{t}dt'
e^{i\omega_{nm}(\bm k)t'}
\braket{n\mathbf{k}|[\hat{\bm r}(0),\hat \rho_0]|m\bm k}\cdot\bm E(t')
,
\label{rhop}
\end{align}
\end{widetext}
where $\omega_{nm}(\bm k)\equiv\omega_{n}(\bm k)-\omega_{m}(\bm k)$
and $E_n(\bm k)=\hbar\omega_n(\bm k))$ are the unperturbed energy
eigenvalues corresponding to the stationary Schrödinger's equation
$\hat H_0\ket{{n\bm k}} =E_{n}(\bm k)\ket{{n\bm k}}$. Notice that
$\hat\rho_0$ has matrix elements
\begin{equation}
  \label{eq:fermi}
  \braket{n\bm k|\hat\rho_{0}|m\bm k}
=\delta_{nm} f(E_{n}(\bm k)),
\end{equation}
with $f$ the Fermi-Dirac distribution, which at the temperature $T=0$ becomes
\begin{align}\label{fd-1}
f(E_{n}(\bm k))
=\Theta(E_F-E_n(\bm k))\equiv f_n(\bm k)
,
\end{align}
with $E_F$ the Fermi energy of the system and $\Theta$ the unit step
function. This defines the distributions functions $f_n(\bm k)$ in
reciprocal space, one for each band.

It is convenient to represent the position operator
in coordinate space $\hat{\bm r}(0)\to \bm r$, when calculating its
{\em interband} matrix elements
and in reciprocal space $\hat {\bm r}(0)\to i\nabla_{\bm k}$ when
calculating its {\em intraband} matrix elements,
so that following Ref.~\onlinecite{aversaPRB95}, we can readily show that
\begin{equation}\label{conmu2}
\braket{n\bm k|[\hat{\mathbf{r}},\hat{\rho}_0]|m\bm k}
= f_{mn}(\bm{k}) \bm r_{nm}(\bm k)
 + i\delta_{nm}\nabla_{\bm k} f_n(\bm k),
\end{equation}
where $f_{nm}(\bm k)=f_n(\bm k)-f_m(\bm k)$ and $\bm r_{nm}(\bm
k)=\braket{n\bm k|\bm r|m\bm k}$. Notice that $f_{nm}(\bm
  k)=0$ if $n=m$.
  The well-known commutator
  \begin{align}\label{v.1}
    \hat{\bm v}=\hat{\dot{\bm r}}=\frac{1}{i\hbar}[\hat{\bm r},\hat H_0],
  \end{align}
allows us to write
the interband matrix element as
\begin{equation}
\bm r_{nm}(\bm k)=\frac{\bm{v}_{nm}(\bm{k})}{i\omega_{nm}(\bm{k})}
\quad (n\neq m)
,
\label{mv}
\end{equation}
where $\hat{\bm v}$ is the velocity operator related to the momentum
operator by $\hat{\bm p}=m_e\hat{\bm v}$.

To obtain the optical linear response we look for the expectation
value of the macroscopic polarization density $\bm P$, whose
time derivative yields the current density, i.e., the expectation
value of the current operator. Thus,
\begin{equation}\label{eq.Pdot}
\frac{\partial}{\partial t}\bm
P=-\frac{e}{\Omega}\mbox{Tr}\,({\hat\rho^{(1)}}(t)\hat{{\mathbf{v}}}(t)),
\end{equation}
with $\Omega$ the
volume of the unit cell.
Assuming an harmonic perturbation
$\bm E(t)=\bm E(\omega)e^{-i\omega t}$,
we obtain the tensorial relation
\begin{equation}
P^{a}(\omega)=\chi^{ab}(\omega) E^{b}(\omega)
,
\label{pshg}
\end{equation}
where $\chi^{ab}(\omega)$ is the linear
susceptibility response tensor,
where the
superscripts $\mathrm{a,b}$ denote Cartesian components, and we use
Einstein convention for repeated indices.
Using
Eqs.~\eqref{rhop}-\eqref{pshg}, we obtain
\begin{widetext}
\begin{align}\label{ee.1}
\chi^{ab}(\omega)
&=
\frac{ie^2}{\hbar\omega}
\sum_{mn}
\int_{\mathrm{BZ}} \frac{d^3k}{8\pi^3}
v^{a}_{mn}(\bm k)
\left(\frac{
f_{mn}(\bm k)
r^{b}_{nm}(\bm k)
+
i\delta_{nm} \partial_{k^{b}} f_n(\bm k)
}{\omega_{nm}(\mathbf{k})-\omega}
\right)
\nonumber\\
&=
\chi_e^{ab}(\omega)
+
\chi_i^{ab}(\omega)
,
\end{align}
\end{widetext}
where the sums over band indices and over the wavevector correspond to
the trace and the latter was replaced by the usual integral over the
first Brillouin zone (BZ), we defined
$\partial_{k^{b}}=\partial/\partial
k^{b}$ to simplify our notation, and we identified the
interband $\chi_e^{ab}(\omega)$ and intraband
$\chi_i^{ab}(\omega)$ contributions to the
susceptibility as those that contain the  first and second terms in the
numerator of Eq. \eqref{ee.1}, those which contain the factor $f_{nm}(\bm k)$ and
the Kronecker's delta $\delta_{nm}$ respectively. Using
Eq. \eqref{mv} we write  the {\it interband} contribution as
\begin{widetext}
\begin{align}\label{ee.2}
\chi_e^{ab}(\omega)
&=
\frac{e^2}{\hbar\omega}
\sum_{mn}
\int_{\mathrm{BZ}} \frac{d^3k}{8\pi^3}
f_{mn}(\bm k)
\omega_{nm}(\bm k)
\left(\frac{
r^a_{mn}(\bm k)
 r^{b}_{nm}(\bm k)
}{\omega_{nm}(\bm k)-\omega}
\right)
.
\end{align}
\end{widetext}
The {\it intraband} contribution is
\begin{widetext}
\begin{align}\label{ee.3}
\chi_i^{ab}(\omega)
&=
\frac{ie^2}{\hbar\omega}
\sum_{mn}
\int_{\mathrm{BZ}} \frac{d^3 k}{8\pi^3}
v^a_{mn}(\bm k)
\left(\frac{
i\delta_{nm}\partial_{k^{b}} f_n(\bm k)
}{\omega_{nm}(\bm k)-\omega}
\right)
%\nonumber\\
%&=
=
\frac{e^2}{\hbar\omega^2}
\sum_{n}
\int_{\mathrm{BZ}} \frac{d\bm{k}}{8\pi^3}
v^a_{n}(\bm{k})
\partial_{k^{b}} f_n(\bm{k})
,
\end{align}
\end{widetext}
where we denote $v^a_{nn}(\bm{k})$ as $v^a_n(\bm{k})$,
the electron's velocity for the $n$-th band at point
$\bm{k}$ of the Brillouin zone.
We recognize
$\chi_e^{ab}(\omega)$ as the well known linear susceptibility
amply discussed in the scientific literature as well as in textbooks of
physics; in the rest of the article, we only devote our attention  to
$\chi_i^{ab}(\omega)$, that as we see now, leads to a closed
and computationally amenable expression for the Drude
frequency.

We rewrite Eq.~\eqref{ee.3} as
\begin{align}\label{ee.4}
\chi_i^{ab}
(\omega)
&=
-\frac{1}{4\pi\omega^2}
{\mathcal D}^{ab}
,
\end{align}
where we define the {\em Drude tensor} as
\begin{align}\label{ee.5}
{\mathcal D}^{ab}
&=-
\frac{4\pi e^2}{\hbar}
\sum_{n}
\int_{\mathrm{BZ}} \frac{d^3k}{8\pi^3}
v^a_{n}(\bm{k})
\partial_{k^{b}} f_n(\bm{k})
\nonumber\\
&=
\frac{e^2}{2\pi^2}
\sum_{n}
\int_{\mathrm{BZ}} d^3k\,
v^a_{n}(\bm{k})
 v^b_{n}(\bm{k})\delta(E_F-E_n (\bm{k}))
.
\end{align}
Here, we used Eq.~\eqref{fd-1} for $f_n(\bm{k})$, employed the
derivative of the Heaviside step function with respect to its argument,
$d\Theta(x)/dx=\delta(x)$, and identified the velocity of the electron
in the $n$-th band from the dispersion of the corresponding energy
$\nabla_{\bm k} E_n(\bm{k})=\hbar \bm v_{n}(\bm k)$. \hl{Notice that the
expression in Eq. (6) of Ref.~\onlinecite{maksimovJPFMP} corresponds to
$\mbox{Tr}\,{\bm{\mathcal D}}/3$ , as it relates to an
isotropic system.}

For any two scalar valued functions $F(\bm k)$ and $G(\bm k)$ of the crystal momentum $\bm k$,
we can evaluate\cite{hormander83}
\begin{align}\label{pze.1}
\int_{\mathrm{BZ}} d^3k\,
F(\bm k)\delta\left(G(\bm k)\right)
 =\int_S d^2\sigma_{\bm k}\,
\frac{F(\bm k)}{|\nabla_{\bm k} G(\bm k)|}
,
\end{align}
which takes us from an integration over the first Brillouin zone with
a Dirac's delta function $\delta(G(\bm k))$ to a surface
integral over that surface $S$ within the first Brillouin zone for which $G(\bm k)=0$, with
$d^2\sigma_{\bm k}$ the differential element of area in reciprocal
space. Notice that $S$ may have zero, one or more connected
components, so that we interpret the surface integral as implying a
sum over all of them.
Applying this result to Eq. \eqref{ee.5} we obtain
\begin{align}\label{drude.2}
{\mathcal D}^{ab}
&=
\frac{e^2}{2\pi^2}
s\sum_{n}
\int_{S_n} d^2\sigma_{\bm k}
\frac{v^{a}_{n}(\bm k)v^b_{n}(\bm k)}{|\nabla_{\bm
                                      k} (E_F-E _n (\bm k))|}
\nonumber\\
&=
\frac{e^2}{2\pi^2\hbar}
s\sum_{n}
\int_{S_n} d^2\sigma_{\bm k}
\frac{v^{a}_{n}(\bm k) v^b_{n}(\bm k)}{v_n(\bm k)}
,
\end{align}
where $S_n$ is the contribution of the $n$-th band to the Fermi Surface
(FS), defined by those Bloch vectors $\bm k$
for which $E_n(\bm k)=E_F$. This is the main result of our paper, as
it allows the calculation of the Drude tensor, the intraband
susceptibility and its contribution to the dielectric response for any
metal from its electronic structure. Notice that $\mathcal D^{ab}$ is
proportional to the integral over the Fermi surface of \hl{the convective
contribution to the momentum flux tensor $m\bm v_n\bm v_n$, with
components $m v_n^a v_n^b$ divided by the
magnitude of the velocity $v_n$}, summed over the conduction bands and integrated over the
Fermi surface. We have included in Eq. \eqref{drude.2}
a spin-degeneracy factor $s$ which
allows us to ignore the spin in the state labels $n$ for those systems
that are spin-degenerate. Thus, we take $s=1$ if
the band index includes the spin, as would be the case when
the spin-orbit coupling is included in the
calculation, and $s=2$ when the spin is not explicitly
included.

For a semiconductor there are no bands that cross the Fermi
Energy, and then the Drude tensor is identically zero, or in other
words, there is no intraband contribution to the susceptibility. For a
metal, there is at least one band that crosses the $E_F$, yielding a
finite
intraband contribution.
The intraband dielectric tensor may be written written as
\begin{align}\label{eps.1}
\epsilon_i^{ab}=\delta_{ab}+4\pi\chi_i^{ab}
=\delta_{ab}-\frac{{\mathcal D}^{ab}}{\omega^2}
,
\end{align}
giving the
characteristic $1/\omega^2$ divergence as $\omega\to 0$ for the dielectric
function of metals. Notice that ${\mathcal D}^{ab}$ has units of
frequency squared and that it would agree with Eq. \eqref{ds.1} for a
Drude model if we identify
$\mathcal D^{ab} =
\omega_D^2\delta_{ab}$ with $\omega_D$ given by Eq. \eqref{ds.2}. Thus, the interband
contribution to the dielectric response given by
Eqs. \eqref{drude.2} and \eqref{eps.1} can be seen a generalization of the Drude model.
\hl{However, we should remark that there are further screening processes
that may shift the actual plasma frequencies away from the Drude
frequency, as illustrated below; the plasma frequencies are given by
the singluarities of the full dielectric response, and not only of its
Drude contribution.
Furthermore,
our dielectric response has an explicit tensorial character} in
contrast to that found in standard textbooks, which always
describe a scalar dielectric response. The symmetry of the system
determines which components of $\mathcal
D^{ab}$ are \hl{non-null} and how they relate among themselves.

We may write the Drude tensor in a more familiar way by
integrating by parts the first line of
Eq.~\eqref{ee.5},
to obtain
\begin{align}\label{ee.10}
\mathcal D^{ab}
&=
\frac{4\pi e^2}{\hbar}
\sum_{n}
\int_{\mathrm{BZ}} \frac{d^3k}{8\pi^3}
f_{n}(\bm{k})
\partial_{k^{b}}
v^a_n(\bm{k})
,
\end{align}
and identify
\begin{align}\label{pz.10}
\partial_{k^{b}}v^{a}_{n}(\bm{k})
%=
%\frac{1}{m_e}\delta_{ab}
%-
%\sum_{m\ne n}
%\frac{v^a_{nm}(\bm{k}) v^{b}_{mn}(\bm{k})+v^a_{mn}(\bm{k}) v^{b}_{nm}(\bm{k})}{\hbar\omega_{mn}(\bm{k%})}
%\equiv
  =
\hbar (m^*_n(\bm{k}))^{-1}_{ab}
,
\end{align}
from Ref.~\onlinecite{cabellosPRB09a},
 where $\bm m^*_n(\bm{k})$ is the effective mass tensor. Then
\begin{widetext}
\begin{align}\label{pz.12}
\mathcal D^{ab}
&=
4\pi e^2
\sum_{n}
\int_{\mathrm{BZ}} \frac{d^3k}{8\pi^3}
f_{n}(\bm{k})
({m^*_n(\bm{k})})^{-1}_{ab}=4\pi n_e e^2
                  \braket{(m^*)^{-1}_{ab}}
,
\end{align}
\end{widetext}
where the sum may be restricted to the conduction bands. This result
is similar to the usual Drude formula with $n_e$ the number density of conduction
electrons but with the inverse electronic mass $1/m_e$ replaced by
an average $\braket{(\bm m*)^{-1}}$ of the inverse effective mass tensor over the occupied
states of the conduction bands.
Although this result looks appealing, it hides the fact that only the
electrons at the Fermi surface do contribute to the dynamics and the
response of the system. Furthermore, the numerical integration over
the full BZ in Eq. \eqref{pz.12} would require
a larger grid  $\{\bm{k}_\alpha\}$ of k-points and would be more
costly to evaluate than the numerical integral \eqref{drude.2}
only over those $\bm k$ points close to the Fermi
surface (see Sec. \ref{si}). Finally, evaluation of the inverse mass tensor in
Eq. \eqref{pz.12} requires a
larger computational cost and has a larger numerical uncertainty than the
evaluation of the velocity matrix elements in Eq. \eqref{drude.2}.

\subsection{Ideal Metal}\label{im}

Within the Sommerfeld theory of metals the conduction electrons have the
dispersion relation of free electrons $E=\hbar^2k^2/2m_e$ and fill up a
Fermi sphere of radius $k_F$, related to the number density $n_e$ of
conduction electrons
through\cite{ashcroft_solid_1976} $n_e=k_F^3/3\pi^2$. \hl{The Fermi
velocity is $v_F=\hbar k_F/m_e$.}
Due to the spherical symmetry, ${\mathcal D}^{ab}=0$
for $a\neq b$, and
${\mathcal D}^{xx}={\mathcal D}^{yy}={\mathcal
  D}^{zz}=\mathcal D^{aa}/3\equiv
\omega_D^2$,
Then, from Eq.~\eqref{drude.2}
\begin{equation}\label{drude.7}
  \omega^2_D
  =\frac{1}{3} \frac{e^2}{2\pi^2\hbar} s \int k_F^2
  d^2\Omega\, v_F = 4\pi \frac{k_F^3}{3\pi^2} \frac{e^2}{m_e}=\frac{4\pi n_e e^2}{m_e},
\end{equation}
where we wrote $d^2\sigma_{\bm k}=k_F^2d^2\Omega$, with $d^2\Omega$
the differential element of solid angle. Thus $\omega_D$ is the
well known Drude value for the plasma frequency, however
derived from
Eq.~\eqref{drude.2} which comes
from a purely quantum mechanical approach within time-dependent
perturbation theory, instead of the text-book derivation from the
phenomenological model of Drude. Curiously, $\omega_D^2$ is
proportional to the number density $n_e$ of conduction electrons, although only
those electrons at the Fermi surface contribute to the integral in Eq. \eqref{drude.2}.

\subsection{Realistic metal}\label{si}

Eq. \eqref{drude.2} is an elegant expression of the Drude tensor
in terms of an integral over the Fermi surface of a tensor formed by
\hl{products of components of the velocity divided by its magnitude $v_n^a
v_n^b/v_n$.} However, for actual applications to realistic models it is
convenient to return to a volume integral. To that end, we realize
that for any function $g(\bm k)$ we can write
\begin{equation}
  \label{eq:bulkvssurf}
  \int_{S_n}d^2\sigma_{\bm k}\,  g(\bm k)=\int_{\mathrm{BZ}} d^3k\, g(\bm k)|\nabla
  f_n(\bm k)|,
\end{equation}
as can be immediately verified using Eqs. \eqref{fd-1} and
\eqref{pze.1}. Thus, we write Eq. \eqref{drude.2} as
\begin{equation}\label{drude.vol}
{\mathcal D}^{ab}
=
\frac{e^2}{2\pi^2\hbar} s
\sum_{n}
\int_{\mathrm{BZ}}
d^3k |\nabla_{\bm k} f_n(\bm{k})|
\frac{v^{a}_{n}(\bm{k}) v^b_{n}(\bm{k})}{v_n(\bm{k})}.
\end{equation}

To carry out the integration numerically, we generate a regular grid $\{\bm
k_\alpha\}$ of points inside the Irreducible Brillouin Zone that corresponds to the
crystallographic group of the metal under study, numbered by a
discrete set of indices $\alpha$. Then, we approximate
$\nabla_{\bm k}f_n(\bm k)|_{\bm k_\alpha}$ through a finite difference
approximation. As the resulting numerical gradient of the Fermi-Dirac
distribution would be null away from the Fermi surface, we can refine
our grid in its vicinity without a substantial increase in the computational cost, where
$E_F$ is obtained by using the prescription
given in Ref.~\onlinecite{burdickPR63}. For those points where the numerical
gradient is non-null we evaluate the components of the velocity. Then
we can replace the integral by a Riemann sum and write
\begin{equation}\label{drude.4nn}
{\mathcal D}^{ab}
=
\frac{e^2}{2\pi^2\hbar} s(\Delta k)^3
\sum_{n}
\sum_\alpha
|\nabla f_n(\bm k_\alpha)|
\frac{v^{a}_{n}(\bm k_\alpha) v^b_{n}(\bm k_\alpha)}{v_n(\bm k_\alpha)},
\end{equation}
where $\Delta k$ is the distance between neighbor vectors $\bm
k_\alpha$ in our grid.

\section{Results}\label{res}

We present results for the three noble metals, Ag, Au and Cu. However,
we do so with more detail for Ag, for which there is a recent accurate experimental
measurement for the Drude frequency
$\hbar\omega_D=8.9\pm 0.2$\,eV.\cite{yangPRB15}
Furthermore, we also consider the case of a metal under applied
stresses consisting of
an isotropic and an anisotropic uniaxial strain.
For the isotropic deformation we simply modify the lattice constant
$a=(1+\gamma) a_0$ with respect to the equilibrium lattice constant
$a_0$ in the absence of stress. For the
anisotropic deformation we keep the original volume of the FCC unit cell,
and deform it by stretching along the $z$ direction
$a_z=(1+\gamma) a_0$ while shrinking it along the $x$ and
$y$ directions, $a_x=a_y=a_\parallel=a_0/\sqrt{1+\gamma}$, thus
converting the FCC-lattice into a BCT-lattice. We allow the
expansion factor $\gamma$ to take negative as well as positive
values. Given the symmetry of these systems, we expect
$D^{ab}=0$ if $a\ne b$, $D^{xx}=D^{yy}\equiv D^\parallel$ in the uniaxial case and
$D^{ab}\equiv D^{\mathrm{iso}}\delta_{ab}$ in the isotropic case.
Therefore, we define the Drude frequencies
\begin{equation}\label{drude.w}
  \begin{aligned}
    \omega^{\mathrm{iso}}_D&=\sqrt{\mathcal D^{\mathrm{iso}}}&\quad\mathrm{isotropic},
    \\
    \omega^{\parallel}_D&=\sqrt{{\mathcal D}^\parallel}&\quad\mathrm{anisotropic},
    \\
    \omega^z_D&=\sqrt{{\mathcal D}^z}&\quad\mathrm{anisotropic}
    .
  \end{aligned}
\end{equation}

The self-consistent ground state and the Kohn-Sham states were
calculated in the DFT-LDA framework using the plane-wave
ABINIT code.\cite{abinit}
We used Troullier-Martins pseudopotentials\cite{troullierPRB91} that are
fully separable nonlocal pseudopotentials in the Kleinman-Bylander
form.\cite{kleinmanPRL82}
We use
$a_0=4.0853$\,\AA, for Ag,
$a_0=3.6149$\,\AA, for Cu,
and
$a_0=4.0782$\,\AA, for Au,
all
taken from Ref.~\onlinecite{ucell}.
For the noble metals only the 6th-band is partially filled and thus crossed by
the Fermi Energy, therefore in
Eq.~\eqref{drude.4nn} the sum over $n$ picks up only the value $n=6$.
Finally, the number of $\bm{k}$-points used for the calculation was
around $\sim 100,000$.

\subsection{Ag}

\begin{figure}[]
\centering
\includegraphics[scale=.65]{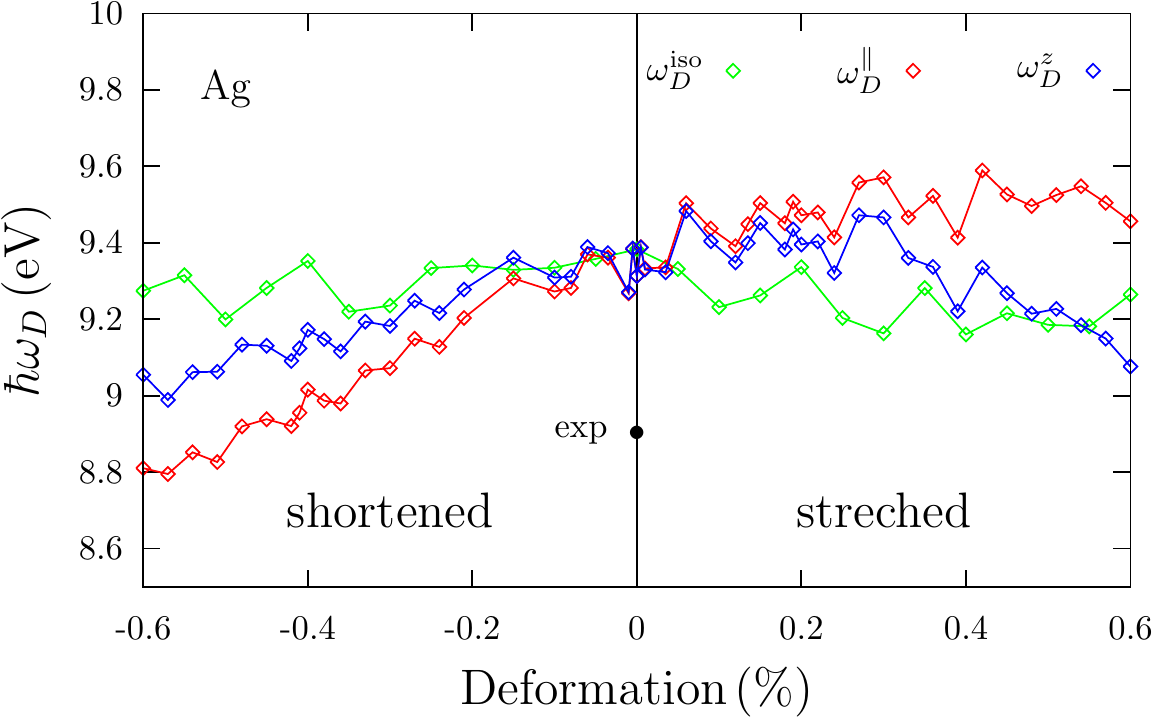}
\caption{(Color Online)  Drude $\hbar\omega_D$ vs. the deformation of
  the unit cell for Ag.
}\label{fig-1}
\end{figure}

In Fig.~\ref{fig-1} we show $\omega^{\mathrm{iso}}_D(\gamma)$, $\omega^\parallel_D(\gamma)$ and
$\omega^z_D(\gamma)$ as a function of the corresponding deformation
$\gamma$ of the unit cell. First, we point out that the three
calculations agree among themselves in the limit of no deformation $\gamma=0$,
and that the resulting value $\hbar\omega_D(\gamma=0)=9.38\mathrm{eV}$
is only ~5\% off from the experimental value $\hbar\omega_D=8.9\pm
0.2\mathrm{eV}$ \cite{yangPRB15}.
The values  chosen for the compressive ($\gamma<0$) and expansive
($\gamma>0$) deformations are experimentally feasible,\cite{akahamaJAP04}
and although small, show a sizable difference
for the anisotropic
deformation values of $\omega^\parallel_D$ and $\omega^z_D$. However,
for the isotropic deformation
$\omega^{\mathrm iso}_D$
does not deviate that much from its undeformed value.

We notice that for very small values of $\gamma$
both isotropic and anisotropic results are almost the same, but as
$\gamma$ deviates away from 0, sizable changes are seen.
The isotropic result $\omega^{\mathrm{iso}}_D(\gamma)$ is relatively
flat with values between 9.2 and 9.4 eV.
On the other hand, the anisotropic
deformation gives an $\omega^{\parallel,z}_D(\gamma)$ that decreases
as $a_z$ is shortened.
Also, as $a_z$ is stretched,
for values of $\gamma\leq 0.3\%$, both
 $\omega^{\parallel}_D(\gamma)$
and  $\omega^z_D(\gamma)$ increase and show similar values,
whereas for larger values of $\gamma\geq 0.3\%$,
$\omega^{\parallel}_D(\gamma)$ continues increasing
while  $\omega^z_D(\gamma)$ reaches a maximum and starts decreasing.
Our results show some oscillations that are due to our
approximating a surface integral through a volume integral represented
as a sum over a discrete grid. As we deform the system, the true Fermi
surface sweeps across the grid points, giving rise to the
oscillations which may be interpreted as indicative of the accuracy of
our calculation. We expect that they could be somewhat filtered away
by approximating the gradient with a higher order finite differences
formula.
It is worth mentioning that, as explained in
Ref.~\onlinecite{yangPRB15}, the experimental error in $\hbar\omega_D$
of $\sim 0.2$\,eV will make our predictions easily
verifiable.

\begin{figure}[]
\centering
\includegraphics[scale=.65]{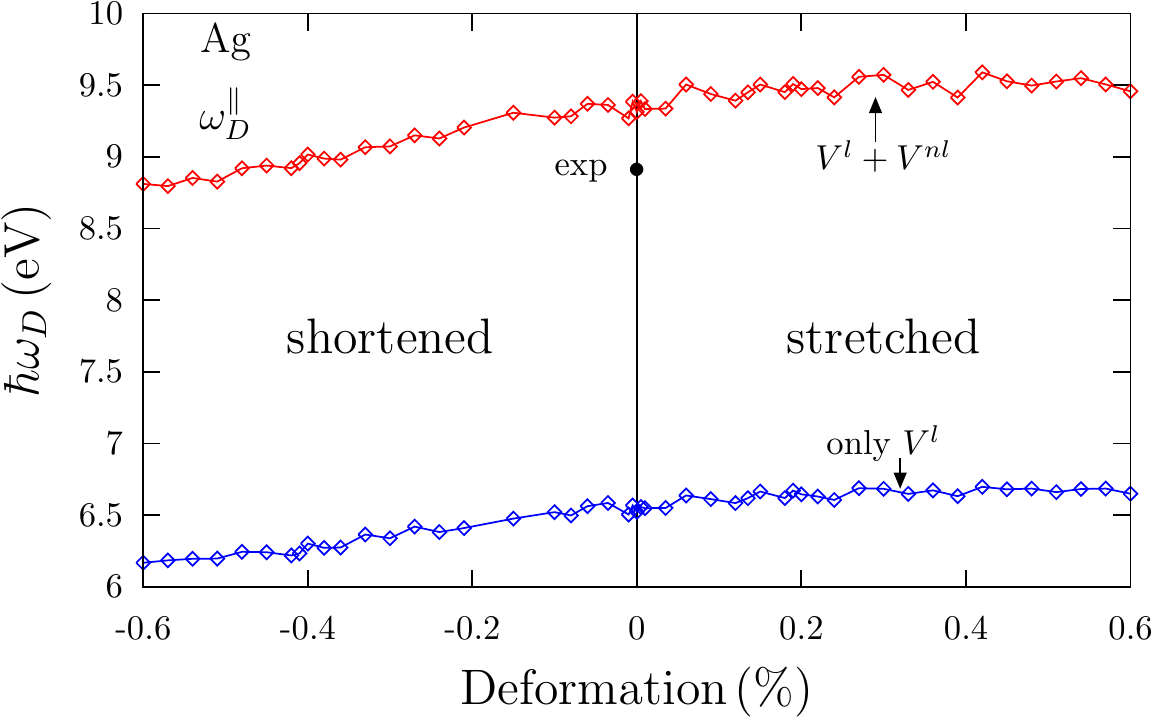}
\caption{(Color Online)  Drude $\hbar\omega^\parallel_D$ vs. the deformation
$\gamma$ of  the unit cell for Ag with and without the nonlocal
potential $\hat{V}^{nl}(\bm{r},\bm{r}')$.
}\label{fig-2}
\end{figure}

An important theoretical point in our formulation is the fact that the
expression for the Drude tensor, Eq.~\eqref{drude.4nn}, depends on
matrix elements of the
velocity operator which we calculated according to Eq. \eqref{v.1}.
We used a non local unperturbed Hamiltonian
\begin{align}\label{v.2}
\hat H_0
=\frac{\hat{p}^2}{2m} + \hat V^{\mathrm{l}} +\hat V^{\mathrm{nl}}
,
\end{align}
where
$\hat{V}^{\mathrm{l}}$ is the local potential characterized by the function
$V^{\mathrm{l}}(\bm r)$, and $\hat{V}^{\mathrm{nl}}$ is the nonlocal
potential characterized by the kernel $V^{\mathrm{nl}}(\bm
r, \bm r')$.
The Schrödinger equation reads
\begin{widetext}
\begin{align}\label{ache.4}
\left(
-\frac{\hbar^2}{2m}\nabla^2
 + V^{\mathrm{l}}(\bm r)\right)\psi_{n\bm k}(\bm r)
 + \int d^3 r'\, \hat{V}^{\mathrm{nl}}(\bm{r},\bm{r}')\psi_{n\bm{k}}(\bm{r}')
=E_n(\bm{k})\psi_{n\bm{k}}(\bm{r})
,
\end{align}
\end{widetext}
The nonlocal potential is more important for metals, where the orbitals of the
conduction electrons are farther away form the nucleus, than for
semiconductors, for which they are usually much closer.
Therefore, we show that it is very important for our calculation
to include the nonlocal contribution to the velocity $\bm{v}_n(\bm{k})$. In  our case,
this was carried out using the DP code.\cite{olevanoDP}
In Fig.~\ref{fig-2} we show $\omega^\parallel_D(\gamma)$, with and
without the contribution of
the nonlocal potential $V^{\mathrm{nl}}(\bm{r},\bm{r}')$. We see that
it is mandatory to include it, otherwise the value of $\omega^\|_D$ would be heavily
underestimated. Similar results are found for $\omega^{\mathrm{iso}}_D$
  and $\omega^z_D$. We remark that the non-locality of the Hamiltonian
arises from its being an effective one particle effective Hamiltonian
for a many-body system.\cite{pulciPRB98}

\begin{figure}[]
\centering
\includegraphics[scale=.8]{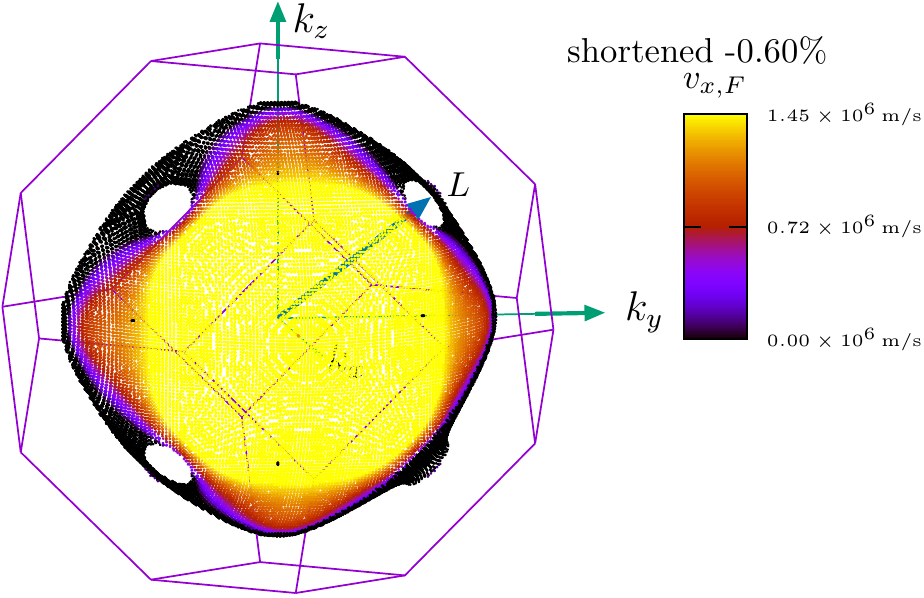}
\includegraphics[scale=.8]{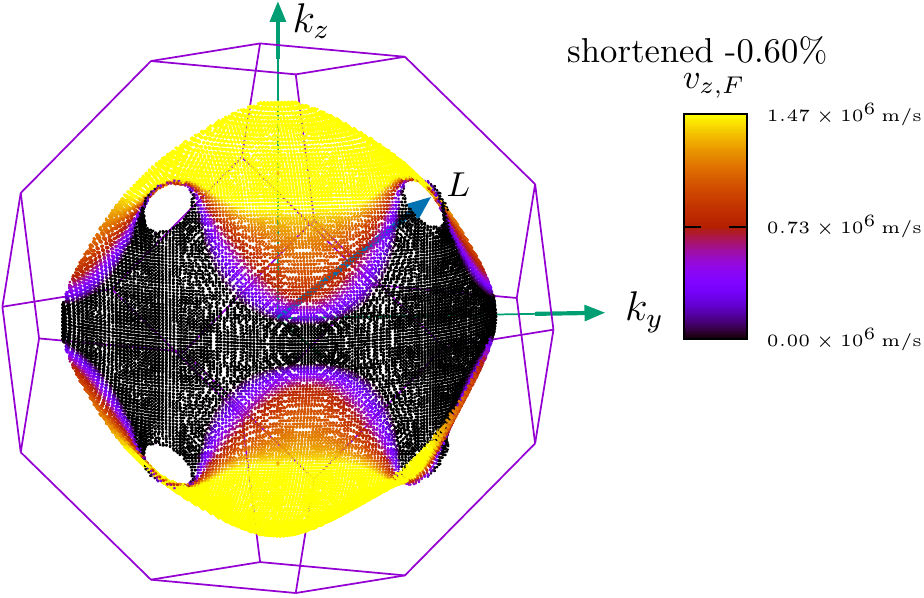}
\includegraphics[scale=.8]{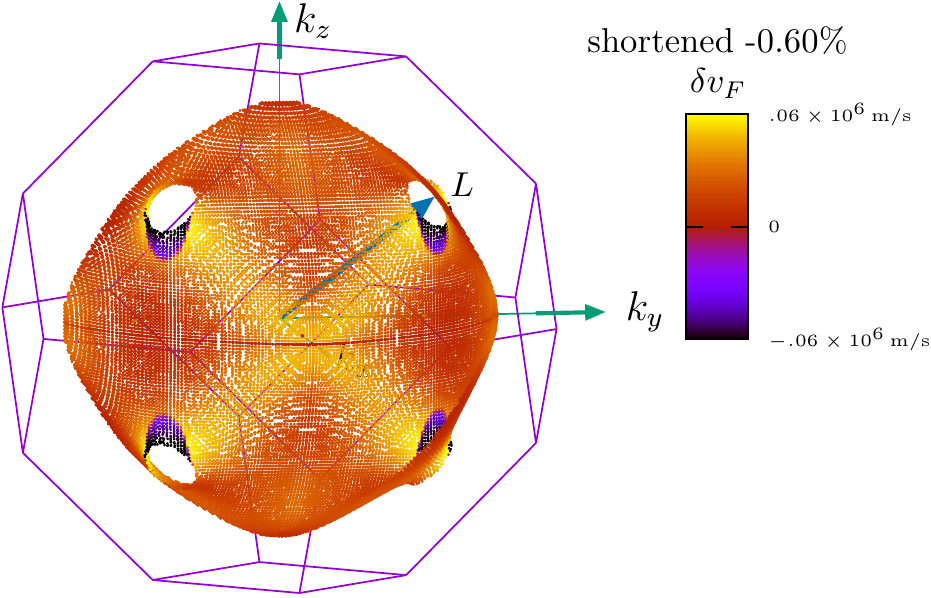}
\caption{(Color Online)
$|v_{x,F}|$ (top-left),
$|v_{z,F}|$ (top-right)
and
$\delta v_F=v_{z,F}(k_x, k_y, k_z)-v_{x,F}(k_z, k_y, k_x)$ (bottom)
 at the Fermi Surface for Ag with  an anisotropic
  deformation of
$\gamma=-0.6\%$.
The necks are centered at the $L$ point.
See the text for details.
}\label{fig-3}
\end{figure}

As explained in Sec.~\ref{si}, the numerical calculation of
Eq.~\eqref{drude.4nn} requires the $\bm{k}_\alpha$-points at the FS as
well as the velocities evaluated at each of these points. In
Fig.~\ref{fig-3} we show
the FS and the Fermi velocity for an anisotropic deformation $\gamma=-0.6\%$
as a illustrative example.
We see that
\hl{the velocities are of the order expected from the Sommerfeld model},\cite{ashcroft_solid_1976}
$\sim 10^6$\,m/s.  The system is symmetric under
reflections on the Cartesian planes and under an inversion around the
origin. As the unit cell has
the same lattice parameter along $x$ and $y$, that $v^x_F$ and $v^y_F$ agree after
a rotation of the system by 90 degrees in the plane, equivalent to an
exchange $k_x \leftrightarrow k_y$, but they differ from $v^z_F$ after
a rotation by 90 degrees around the $y$ axis due
to the uniaxial strain. Fig. \ref{fig-3} shows that in
this case the $x-z$ anisotropy $\delta
v_F=v^z_F(k_x,k_y,k_z)-v^x_F(k_z,k_y,k_x)$ is of the order of a few
percent.
It is quite interesting to see that around the FS necks,
centered at the $L$-point the velocity is not uniform.

\begin{figure}
  \centering
\includegraphics[scale=.8]{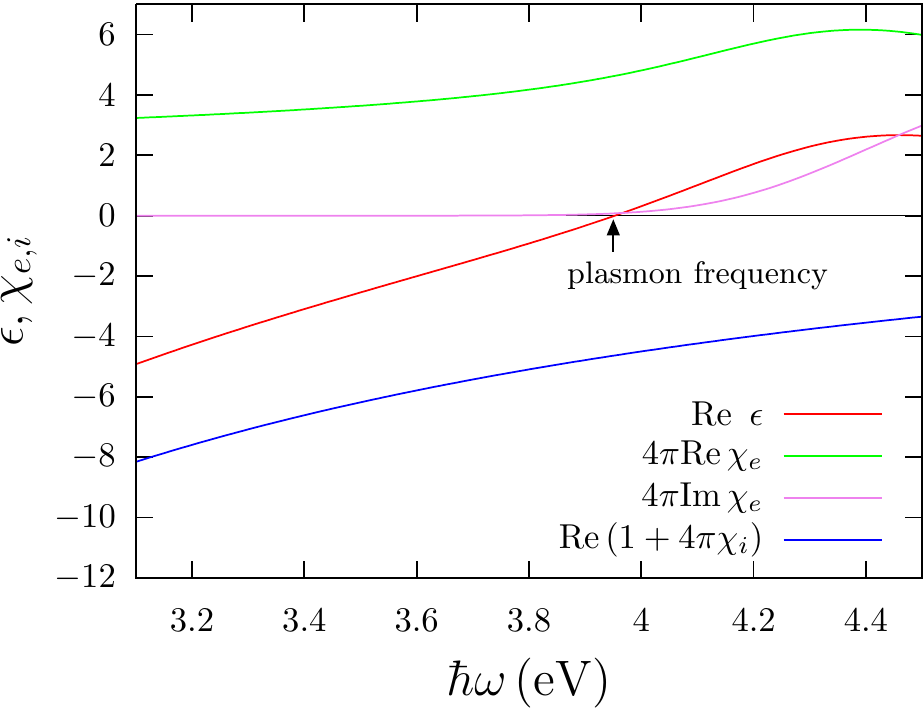}
  \caption{\hl{Full dielectric function of unstrained Ag and its interband
    and intraband contributions. Notice that the plasma frequency,
    given by its zero, differs significantly from the Drude
    frequency.}}
  \label{fig:fullAg}
\end{figure}

\hl{In order to illustrate the difference  remarked previously between
the Drude and the plasma frequencies, in Fig. \ref{fig:fullAg} we show
the dielectric function of unstrained Ag incorporating both its interband and its
intraband contributions. The intraband contribution has a peak due to
d-sp transitions below the Drude frequency, and this pulls the
dielectric function upwards, which thus, crosses zero at the plasmon energy
$\hbar\omega_p=3.95\,\mbox{eV}/\hbar$,
much lower than the Drude result $\hbar\omega_D=9.36\,\mbox{eV}$. To
obtain this result we accounted for the difference in energy of the
excited bands as compared to the DFT-LDA prediction, by using the GW
formalism within the scissors operator approximation, as
described in Ref.~\onlinecite{cabellosPRB09a}.}

\subsection{Cu and Au}

\begin{figure}[]
\centering
\includegraphics[scale=.65]{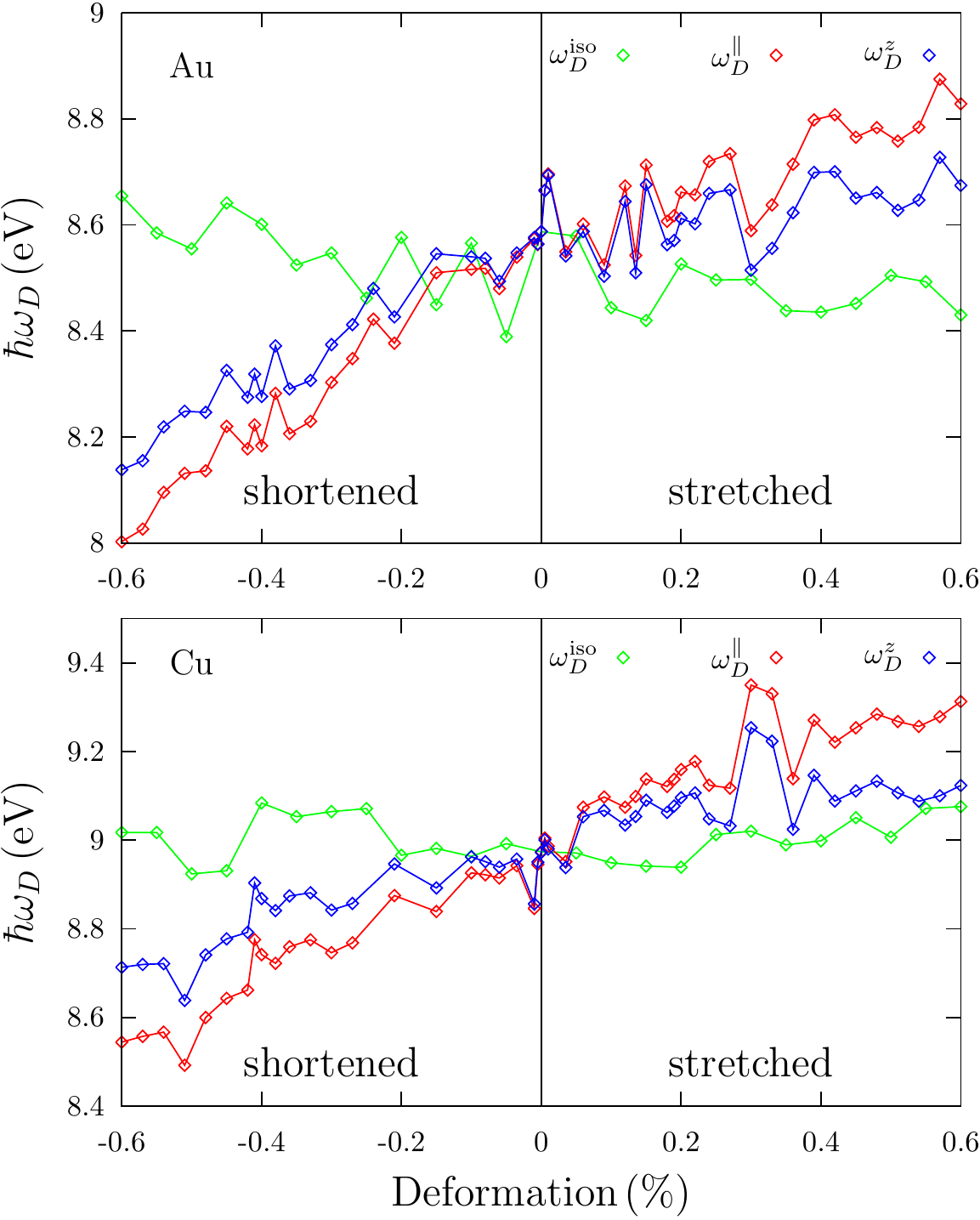}
\caption{(Color Online)  Drude $\hbar\omega_D$
 vs. the deformation of
  the unit cell for Au and Cu.
}\label{fig-5}
\end{figure}

In Fig.~\ref{fig-5} we show $\omega^{\mathrm{iso}}_D(\gamma)$, $\omega^\parallel_D(\gamma)$ and
$\omega^z_D(\gamma)$  as a function of the
corresponding deformation $\gamma$ of the
unit cell for Cu and Au.
As for Ag,
we notice that for very small values of $\gamma$
both isotropic and anisotropic results are almost the same, but as
$\gamma$ deviates away from 0, sizable changes are seen. The
qualitative behavior
of $\omega_D$ for Cu and Au is very similar to that of Ag, and thus
we do not describe it again for briefness sake.
The
theoretical values
of
$\hbar\omega_D(\gamma=0)=8.97$\,eV for Cu and
$\hbar\omega_D(\gamma=0)=8.59$\,eV for Au for the unstrained case
are close to the experimental results reported in
Ref.~\onlinecite{plasma} and shown in Table.~\eqref{tab:1}.
However, the experimental values, except that of
Ag
in Ref.~\onlinecite{yangPRB15}, have been reported without stating the experimental
uncertainty, and there are some noticeable  differences among various
experimental results, so there is definitely a need for more precise
experimental measurements of the Drude plasma frequency for Cu and
Au. However, we point out that our results are
well within the dispersion of the available experiments.

\begin{figure}
  \centering
\includegraphics[scale=.65]{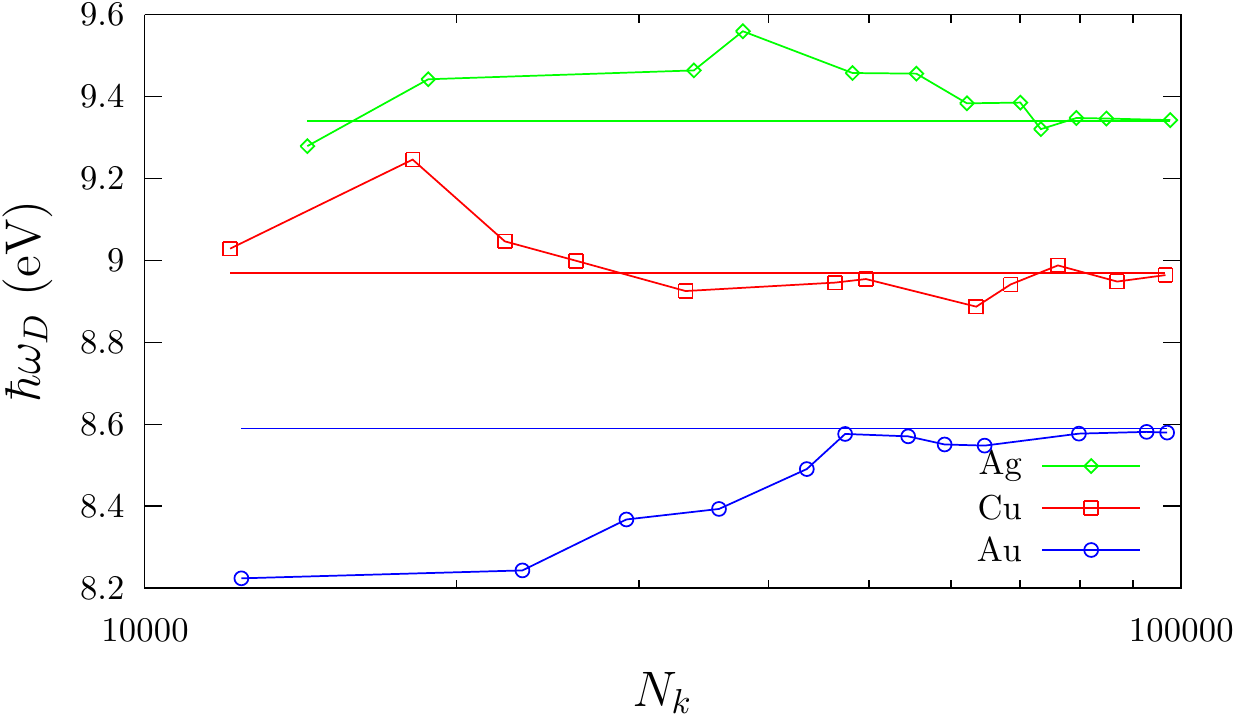}
  \caption{\hl{Drude $\hbar\omega_D$  of unstrained Ag, Cu and Au as a function
    of the number of grid points in the vicinity of the Fermi surface
    employed in its calculation.}}
  \label{fig:convergencia}
\end{figure}

\hl{Finally, in Fig. \ref{fig:convergencia} we illustrate the convergence
of our results as the number of grid points in the vicinity of the
Fermi surface is increased. It may be
appreciated that our results are adequately converged.}

\subsection{Previous Work}

In
Ref.~ \onlinecite{mariniPRB01}, the
Drude plasma frequency was obtained through an expression derived from
the intraband part of the dielectric function,
in the same spirit as
we have done here,
but using a non-local longitudinal susceptibility that depends on the value
of a wavevector $\bm q$, for which the limit $q\to 0$
had to be numerically approximated, being fixed at a rather arbitrary chosen
small value
$|\bm{q}|=0.005$\,a.u.
Furthermore, a fictitious finite electronic
temperature $T$ was used to smooth out the Fermi-Dirac distribution function. As we
show in appendix \ref{compa}, taking the actual limits $q\to 0$ and $T\to 0$ of their
expression, we recover our Eq.~\eqref{ee.5}. However, in their
approach, different
directions of $\bm q$ should be used to obtain all the
components of the tensor $\mathcal D^{ab}$ while our approach does not
require extraneous parameters. The results of
Refs.~\onlinecite{mariniPRB01} and \onlinecite{mariniPRB02},  give
$\hbar\omega_D=9.27$ for Cu and
$\hbar\omega_D=9.48$\,eV for Ag, respectively, that are not far from our
values, as shown in Table ~\eqref{tab:1}.

Also,
in
Ref.~\onlinecite{maksimovJPFMP},
using a similar procedure to that of Ref.~\onlinecite{mariniPRB01},
the
Drude plasma frequency
was obtained
by taking the limit $q\to 0$ analytically.
However, the resulting expression
is valid  only for  isotropic systems, for which it agrees with
$\mathcal D^{\mathrm{iso}}$ as derived from
our  Eq.~\eqref{ee.5}.
Finally, the expression for $\mathcal D^{\mathrm{iso}}$, as derived in
Ref.~\onlinecite{maksimovJPFMP}, was
used in Ref.~\onlinecite{pedersenJPCM09} for the study of  tin and
in Ref.~\onlinecite{jungJAP13}  the study
of of heavily doped semiconductors.

\begin{widetext}
\begin{center}
\begin{table}[t!]
\begin{tabular}{ccccc} \hline
  & Experimental&\multicolumn{2}{c}{Theoretical}\\
  && This work  & Others\\\hline
Ag&8.9$\pm$0.2\cite{yangPRB15}, 8.6\cite{hooperPRB02},
    9.013\cite{ordalAO85}, 9.04\cite{zemanJPC87},
    9.6\cite{blaberJPC09}& 9.38 & 9.48\cite{mariniPRB02} \\
Cu & 7.389\cite{ordalAO85}, 8.76\cite{zemanJPC87} & 8.97 & 9.27\cite{mariniPRB01}\\
Au & 7.9\cite{kreiterPRB02}, 8.55\cite{blaberJPC09},
     8.89\cite{zemanJPC87}, 9\cite{berciaudNL05},
     8.951\cite{gradyCPL04}, 9.026\cite{ordalAO85} &8.59 & -\\
\end{tabular}
\caption{Experimental and theoretical values of $\hbar\omega_D$ in eV
  for Ag, Cu, and Au in the absence of strain.}
\label{tab:1}
\end{table}
\end{center}
\end{widetext}

\hl{Finally, in Ref. \onlinecite{cazzanigaPRB20103} the
reciprocal-vector-dependent intraband contribution to the dielectric
response was obtained through an analytical expansion of several terms
to low order in the small optical wavenumber. Their expressions
could be used to include local field corrections, and they
were applied to RPA-like calculations of the response of Fe and Mg,
obtainaining good agreement with experiment. They emphasized, as we
did above,  the
importance of using non-local pseudopotentials. Their formalism for
the dielectric function is similar to ours, though
we did not require an expansion for the long wavelength case and they
did not identify explicitly an expression for the Drude tensor.}

\section{Conclusions}\label{conc}

We have shown that the well known Drude plasma frequency,  $\omega_D^2$,
should be replaced by the  Drude tensor $\bm{\mathcal D}$,
for which
we have derived a closed theoretical expression
based on
quantum mechanical time dependent perturbation theory.
The expression for
$\bm{\mathcal D}$ as the integral over the Fermi surface of a tensor
built up from matrix elements of the velocity operator
is amiable to analytic and numerical
evaluation.
Using
the Sommerfeld model for metals, we showed that
$\bm{\mathcal D}$
leads to the well known result of
$\omega^2_D=4\pi n e^2/m_e$.
We calculated
$\bm{\mathcal D}$ for the noble metals, and we described the results
for Ag
in depth, since there is a recent precise result of $\omega_D=8.9\pm 0.2$\,eV.\cite{yangPRB15}
We showed that non-local potentials ought to be used when calculating
the velocity matrix elements in order to obtain an accurate result.
In summary, the results of $\omega_D$ for  Cu, Ag and Au for the undeformed unit cell
with a lattice constant taken from measurements at room temperature,
coincide rather well with the experimental results.
We have made predictions for the noble metals
subject to an isotropic and
to an uniaxial stress
that could be experimentally verified.

\section*{Acknowledgments}
We acknowledge useful discussions with Raksha
Singla.

\section*{Funding}
B.S.M. acknowledges  the support from CONACyT through
grant A1-S-9410. W.L.M. acknowledges  the support from DGAPA-UNAM
under grant IN111119.

\section*{Disclosures}
The authors declare no conflicts of interest.

\appendix

\section{Comparison with Marini et al. \cite{mariniPRB01}}\label{compa}

Eq. (17) of Marini et al.\cite{mariniPRB01}, derived from the local
limit of the longitudinal non-local susceptibility,
reads
\begin{widetext}
\begin{align}\label{zp.2}
\hbar^2\omega^2_D&=\lim_{\bm{q}\to 0}\frac{8\pi e^2}{|\bm{q}|^2}\int\frac{d^3k}{8\pi^3}
\left(f_{n}(\bm{k}-\bm{q})-f_{n}(\bm{k})\right)
\Theta\left(f_{n}(\bm{k}-\bm{q})-f_{n}(\bm{k})\right)
\nonumber\\
&\times
\left|
\langle n\bm{k}|e^{i\bm{q}\cdot\bm{r}}|{n\bm{k}-\bm{q}} \rangle
\right|^2
\left(E_{n}(\bm{k}-\bm{q})-E_{n}(\bm{k})\right)
,
\end{align}
\end{widetext}
where $\bm{q}$ is the wavevector of the field.
We use $\bm{k}\cdot\bm{p}$ theory
to approximate this expression for small $\bm q$ using the following results:
\begin{align}\label{zp.3}
E_{n}(\bm{k}+\bm{q})&=E_{n}(\bm{k})+\hbar\bm{q}\cdot\bm{v}_{nn}(\bm{k})
+{\mathcal O}(q^2)
\nonumber\\
\ket{n\bm{k} +\bm{q}}&=e^{i\bm{q}\cdot\bm{r}}\left(\ket{n\bm
                       {k}}+\ket{n\bm{k}}^{(1)} +{\mathcal O} (q^2)\right)
\nonumber\\
\ket{n \bm{k}}^{(1 )}&=\hbar\sum_{m\ne n}\frac{\bm{q}\cdot\bm
                       v_{mn}(\bm k)}{E_{n}-E_{m}}\ket{m \bm{k}}
.
\end{align}
Then, to first order in $\bm q$, we
obtain,
\begin{align}\label{zp.5nn}
\ket{n\bm{k}+\bm{q}}&=(1+i \bm{q} \cdot \bm{r})\left(\ket{n\bm{k}}
+\ket{n\bm{k}}^{(1)}+{\mathcal O} (q^2)\right)
\nonumber\\
&\approx\ket{n \bm{k}} +i \bm{q} \cdot \bm{r} \ket{n \bm{k}} +\ket{n\bm{k}}^{(1)}
,
\end{align}
\begin{widetext}
\begin{align}\label{zp.4}
\braket{n\bm{k}|e^{i\bm{q}\cdot\bm{r}}|{n\bm{k}-\bm{q}}} &\approx
\bra{n\bm{k}}\left(1+i\bm{q}\cdot\bm{r})(\ket{n \bm{k}} -i \bm{q} \cdot \bm{r}
\ket{n \bm{k}} +\ket{n\bm{k}}^{(1)}\right)
\nonumber\\
&=1+\braket{n\bm k|n\bm k}^{(1)}
\nonumber\\
&=
1+\hbar\sum_{m\ne n}\frac{\bm q\cdot \bm v_{mn}(\bm k)}{E_{n}-E_{m}}
              \braket{n\bm k|m \bm k}
=1+\hbar\sum_{m\ne n}\frac{\bm q\cdot\bm v_{mn}(\bm k)}{E_{n}-E_{m}}\delta_{nm}
=1,
\end{align}
\end{widetext}
% Now,
% \begin{align}\label{zp.7}
% f_{n}(\bm{k}-\bm{q})-f_{n}(\bm{k})\approx-\bm{q}\cdot\nabla_{\bm k} f_{n}(\bm{k})
    %     =\hbar\bm{q}\cdot\bm{v}_{nn}(\bm{k})  \delta(E_F-E _n (\bm{k}))
% \end{align}
and
\begin{align}\label{zp.9}
E_{n}(\bm{k})-E_{n}(\bm{k}-\bm{q})
\approx
\hbar \bm{q}\cdot\bm{v}_{nn}(\bm{k})
.
\end{align}
With these results,  and denoting $\bm{v}_{nn}(\bm k)$ as
$\bm{v}_n(\bm k)$, Eq.~\eqref{zp.2} may be written as
\begin{widetext}
\begin{align}\label{zp.10}
\hbar^2\omega^2_D
&=
\lim_{\bm{q}\to 0}\frac{8\pi e^2}{|\bm{q}|^2}\int\frac{d^3k}{8\pi^3}
\hbar\bm{q}\cdot\bm{v}_n(\bm{k})  \delta(E_F-E _n (\bm{k}))
\Theta\left(f_{n}(\bm{k}-\bm{q})-f_{n}(\bm{k})\right)
\hbar \bm{q}\cdot\bm{v}_n(\bm{k})
\nonumber\\
\omega^2_D&=\frac{e^2}{2\pi^2}
\lim_{\bm{q}\to 0}\frac{1}{|\bm{q}|^2}\int  d^3k
(\bm{q}\cdot\bm{v}_n(\bm{k}))^2
\delta(E_F-E _n (\bm{k}))
,
\end{align}
\end{widetext}
where the factor of 1/2 comes from the step function $\Theta$.
The expression above must be calculated for a given direction of
$\bm{q}$. For example, taking $\bm q$ along $x$, we get
\begin{align}\label{zp.11}
\omega^2_D&=\frac{e^2}{2\pi^2}
\int  d^3k
(v^x_n(\bm{k}))^2
\delta(E_F-E _n (\bm{k}))
,
\end{align}
equivalent to Eq.~\eqref{ee.5} for the component
${\cal D}^{xx}$, a the diagonal term of the Drude tensor. This is
enough for isotropic systems.
To get from Eq.~\eqref{zp.10} the full Drude tensor
for an anisotropic system using this formulation we would have
to repeat the calculation for different directions of $\bm q$.
Also, a numerical implementation of
our approach does not require
a finite value of $\bm{q}$ nor a fictitious finite temperature as
used in Refs.~\onlinecite{mariniPRB02} and \onlinecite{mariniPRB01}.

\bibliographystyle{apsrev}
\bibliography{ref.bib}
\end{document}